\documentclass[a4paper]{article}

\usepackage{graphicx}
\usepackage{bm}
\usepackage{amsmath}
\usepackage{amssymb}
\usepackage{latexsym}
\usepackage{exscale}
\usepackage{geometry}

\begin{document}

\title{Conditions for entanglement in multipartite systems}

\author{Mark Hillery$^{1}$, Ho Trung Dung$^{1,2}$, Hongjun Zheng$^{1}$ \\
$^{(1)}$Department of Physics, Hunter College of CUNY \\ 695 Park Avenue \\ New York, NY 10065 \and
$^{(2)}$ Institute of Physics, Academy of Sciences and Technology \\ 1 Mac Dinh Chi Street,
District 1 \\ Ho Chi Minh City, Vietnam}

\date{\today}
\maketitle

\begin{abstract}
We introduce two entanglement conditions that take the form of inequalities involving expectation
values of operators.  These conditions are sufficient conditions for entanglement, that is if they are
satisfied the state is entangled, but if they are not, one can say nothing about the entanglement
of the state.  These conditions are quite flexible, because the operators in them are not 
specified, and they are particularly useful in detecting multipartite entanglement.  We explore the
range of utility of these conditions by considering a number of examples of entangled states, and
seeing under what conditions entanglement in them can be detected by the inequalities presented
here. 
\end{abstract}

\section{Introduction}

Besides being of fundamental interest, entanglement among more than two parties
can potentially be an important resource in quantum communication
and information processing \cite{Braunstein05,Horodecki09}. Quantum
teleportation, quantum dense coding, quantum telecloning and quantum
key distribution schemes involving two parties are extendible to an
arbitrary number of parties sharing multipartite entanglement.
Further proposals that exploit the multiparty quantum correlations
of multipartite entangled states include quantum secret sharing,
where parties may share quantum information retrievable only when
all parties cooperate \cite{Hillery99}, remote concentration of
quantum information \cite{Murao01}, and measurement-based quantum
computing \cite{Nest06}.

The structure of entanglement in multipartite systems is much richer
than that in the case of bipartite systems. Despite the fact that considerable
effort has been spent on characterizing multipartite entanglement, the
detection, classification, and quantification  of entanglement for
arbitrary states of multipartite systems remains a formidable task
\cite{Braunstein05,Horodecki09}. In this paper we focus on the
problem of  detecting entanglement in multipartite systems using inequalities.
One possible strategy in this approach is to use pairwise
inequalities to check for entanglement in every possible bipartite
cut in the system. In this way one may gain detailed information about
which subsystems are entangled \cite{Hillery06b,Shchukin06}.
However, the amount of work required to perform the task can grow
enormously as the number of subsystems increases. It is desirable
to have multipartite inequalities that would allow one to check for
overall entanglement in multipartite systems in a straightforward
and transparent manner. For systems of 
$n$
qubits, inequalities of 
this type exist \cite{sorensen}-\cite{krammer09}.  These inequalities
typically involve collective spin operators, and are simple to apply. 

We shall present two inequalities in this paper that detect the
presence of entanglement in multipartite systems.  These are an outgrowth
of earlier work on entanglement in continuous-variable systems.  Within the
last few years, several papers have presented inequalities for detecting
entanglement in two-mode continuous-variable systems, which are particularly
useful for non-Gaussian states \cite{Hillery06b,Shchukin06},\cite{Hillery06} - \cite{Li07}.  
We note that the papers \cite{Hillery06b,Shchukin06,Li07} dealt with multipartite entanglement.
The inequalities are sufficient conditions for entanglement, if they are satisfied, the
state is entangled, but if they are not, nothing can be concluded. In most cases
these inequalities can be derived from
the partial transpose condition, though the inequalities in \cite{Hillery06} were
not originally proved in this way (see \cite{Nha06,Hillery09}).  In fact, the inequalities
in \cite{Hillery06} provide sufficient conditions to detect entanglement in any bipartite system,
not just in continuous-variable systems,
and they have been applied to explore entanglement in two-mode field states \cite{Hillery06b},
spin systems \cite{Zeng10}, and atom-field entanglement \cite{Hillery09}.

Let us now state the entanglement conditions for multipartite systems, which are the subject
of this paper.  Suppose we have a system consisting of $n$ subsystems, and let $A_{k}$
be an operator on the Hilbert space of the $k^{\rm th}$ subsystem.  A state is entangled if
either of the two conditions
\begin{eqnarray}
\label{cond1}
     && \Bigl|\Bigl\langle \prod_{k=1}^n A_k \Bigr\rangle \Bigr| >
        \prod_{k=1}^n  \langle (A^\dagger_k A_k)^{n/2}\rangle^{1/n},
\\
\label{cond2}
     && \Bigl|\Bigl\langle \prod_{k=1}^n A_k \Bigr\rangle \Bigr| >
           \Bigl\langle
             \Bigl( \frac{1}{n}\sum_{k=1}^n  A^\dagger_k A_k \Bigr)^{n/2}
             \Bigr\rangle,
\end{eqnarray}
is satisfied.  These inequalities are applicable to systems of continuous-variable type, 
discrete type, or a mixture between the two.  We shall first prove these inequalities, and
then proceed to discuss their consequences by making use of several examples.

\section{Separability Conditions}

Consider a system consisting of $n$ subsystems with Hilbert space
${\cal H} = {\cal H}_1\otimes {\cal H}_2\cdots\otimes {\cal H}_n$.
If the system is in a pure state, it is fully separable if and only
if the state is a product of pure states describing $n$ elementary
subsystems. If the state is mixed, it is fully separable if $\rho$
is a statistical mixture of product states
\begin{equation}
\label{e1}
        \rho =\sum_j p_j\rho_j= \sum_j p_j \rho_j^{(1)} \otimes \rho_j^{(2)}\cdots
           \otimes \rho_j^{(n)}.
\end{equation}
Let $A_k$ be an operator on ${\cal H}_k$, then we have for a fully separable state that
\begin{eqnarray}
\label{e2}
     \Bigl|\Bigl\langle \prod_{k=1}^n A_k \Bigr\rangle \Bigr| && =
     \Bigl|\sum_j p_j  \prod_{k=1}^n \langle A_k \rangle_j \Bigr|
\nonumber\\
      &&\le \sum_j p_j \Bigl| \prod_{k=1}^n \langle A_k \rangle_j \Bigr|
\nonumber\\
      &&\le \sum_j p_j
        \prod_{k=1}^n \langle |A_k|^2 \rangle_j^{1/2} \ .
\end{eqnarray}
where $\langle A_k \rangle_j = {\rm Tr}\bigl (A_k \rho_j \bigr)$, and
$|A_k|$ denotes $\sqrt{A^\dagger_k A_k}$. In the first line we used the full separability of the state and in going from the second line to the third, we
used the fact that any operator has a non-negative variance
$$
|\langle A_k \rangle_j| \le \langle |A_k|^2 \rangle_j^{1/2}.
$$
We prove now a lemma.
\\
{\bf Lemma}: For any positive operator $B$ we have that
$\langle B \rangle ^p \le \langle B^p \rangle, \ p>1$.

{\it Proof}: First we write $\langle B \rangle$ in the form
\begin{equation}
\label{e8}
          \langle B \rangle = \sum_{l=1}^m \lambda_l \langle P_l\rangle,
\end{equation}
where $P_l$ is the projector corresponding to $\lambda_l$ and
$\langle P_l\rangle={\rm Tr}(\rho P_l)$.
We shall make use of the H\"{o}lder inequality \cite{Hardy}, which is
\begin{equation}
\label{e9}
      \sum_{l=1}^m |x_l y_l| \le
       \Bigl( \sum_{l=1}^m |x_l|^p \Bigr)^{1/p}
       \Bigl( \sum_{l=1}^m |y_l|^q \Bigr)^{1/q} ,
\end{equation}
where
\begin{equation}
\label{e10}
      \frac{1}{p}+\frac{1}{q}=1, \qquad p>1,\ q>1,
\end{equation}
and the equality holds iff $|x_1|^{p-1}/|y_1| = |x_2|^{p-1}/|y_2|
= \cdots = |x_m|^{p-1}/|y_m|$. For $p=q=2$ it reduces to the Cauchy-Schwarz
inequality. If we set
\begin{equation}
\label{e11}
      x_l = \lambda_l \langle P_l\rangle^{1/p}, \qquad
      y_l = \langle P_l\rangle^{1/q},
\end{equation}
where $p$ and $q$ satisfy Eq. (\ref{e10}), in the H\"{o}lder
inequality, it follows that
\begin{eqnarray}
\label{e12}
      \sum_{l=1}^m |\lambda_l \langle P_l\rangle |
       = \sum_{l=1}^m \lambda_l \langle P_l\rangle^{1/p}
              \langle P_l\rangle^{1/q}
      \le \Bigl( \sum_{l=1}^m \lambda_l^p \langle P_l\rangle \Bigr)^{1/p}
           \Bigl(\sum_{l=1}^m \langle P_l\rangle\Bigr)^{1/q}
      = \Bigl( \sum_{l=1}^m \lambda_l^p \langle P_l\rangle \Bigr)^{1/p},
\end{eqnarray}
hence
$    \langle B \rangle  \le \langle B^p \rangle^{1/p}$.
\hfill $\blacksquare$

\subsection{Derivation of condition (\ref{cond1})}

We shall employ the generalized H\"{o}lder inequality \cite{Hardy}, which is
\begin{equation}
\label{e16}
     \Bigl( \sum_j p_j a_j^r b_j^r \ldots l_j^r \Bigr)^{\frac{1}{r}} \le
     \Bigl( \sum_j p_j a_j^{r/\alpha} \Bigr)^{\alpha/r}
     \Bigl( \sum_j p_j b_j^{r/\beta} \Bigr)^{\beta/r}
      \ldots
     \Bigl( \sum_j p_j l_j^{r/\gamma} \Bigr)^{\gamma/r}\ ,
\end{equation}
where
\begin{equation}
\label{e17}
      \sum_j p_j = 1,\qquad \alpha+\beta+\ldots+\gamma=1.
\end{equation}
Setting $r=1$, $\alpha=\beta=\ldots=\gamma=\frac{1}{n}$, and
$a_j=\langle |A_1|^2 \rangle_j^{1/2}$, 
$b_j=\langle |A_2|^2 \rangle_j^{1/2},\ldots$, 
$l_j=\langle |A_n|^2 \rangle_j^{1/2}$, 
the inequality (\ref{e16}) readily yields
\begin{eqnarray}
\label{e18}
    \sum_j p_j \prod_{k=1}^n \langle |A_k|^2 \rangle_j^{1/2}
    && \le \prod_{k=1}^n \Bigl( \sum_j p_j
             \langle |A_k|^2\rangle_j^{n/2} \Bigr)^{1/n}
\nonumber\\
    && \le \prod_{k=1}^n \Bigl( \sum_j p_j
             \langle |A_k|^n\rangle_j \Bigr)^{1/n}
\nonumber\\
    &&  = \prod_{k=1}^n  \langle |A_k|^n\rangle^{1/n},
\end{eqnarray}
where in the second step we made use of the lemma. This and Eq.
(\ref{e2}) lead to
\begin{equation}
\label{e19}
     \Bigl|\Bigl\langle \prod_{k=1}^n A_k \Bigr\rangle \Bigr| \le
        \prod_{k=1}^n  \langle (A^\dagger_k A_k)^{n/2}\rangle^{1/n}.
\end{equation}
Since all fully separable states must satisfy the inequality
(\ref{e19}), a state that violates
it is an entangled state and we obtain the multipartite entanglement condition (\ref{cond1}).

\subsection{Derivation of condition (\ref{cond2})}

To derive Eq. (\ref{cond2}) we make use of the fact that the geometric mean is 
smaller than or equal to the arithmetic mean
\begin{equation}
\label{e4}
       \prod_{k=1}^n a_k^{1/n} \le \frac{1}{n} \sum_{k=1}^n a_k, \qquad
       a_k\ge 0,
\end{equation}
with equality holding iff $a_1=a_2=\cdots=a_n$. With 
$a_k=\langle |A_k|^2 \rangle_j^{1/2}$, the inequality (\ref{e4}) yields
\begin{eqnarray}
\label{e5}
      \prod_{k=1}^n \langle |A_k|^2 \rangle_j^{1/2}
      && \le \frac{1}{n^n} 
          \left( \sum_{k=1}^n \langle |A_k|^2 \rangle_j^{1/2} \right)^n
\nonumber\\
      &&  \le
    \frac{1}{n^n} n^{n/2} \biggl(
         \sum_{k=1}^n \langle |A_k|^2 \rangle_j \biggr)^{n/2}
\nonumber\\
      &&        =
    \frac{1}{n^{n/2}}
         \biggl\langle \sum_{k=1}^n  |A_k|^2 \biggr\rangle_j ^{n/2}
\nonumber\\
      &&  \le
    \frac{1}{n^{n/2}}
         \biggl\langle \left( \sum_{k=1}^n  |A_k|^2 \right)^{n/2}
            \biggr\rangle_j \ ,
\end{eqnarray}
where in going from the first line to the second we applied 
the Cauchy-Schwarz inequality and in going from the third line to the fourth,
we used the result of the lemma with 
$          B = \sum_{k=1}^n  |A_k|^2$ and $p=n/2$.
The inequality (\ref{e5}) leads to
\begin{equation}
\label{e14}
      \sum_j p_j \prod_{k=1}^n \langle |A_k|^2 \rangle_j^{1/2}
      \le \frac{1}{n^{n/2}} \sum_j p_j 
         \biggl\langle \left( \sum_{k=1}^n  |A_k|^2 \right)^{n/2}
            \biggr\rangle_j 
      = 
       \frac{1}{n^{n/2}}
         \biggl\langle \left( \sum_{k=1}^n  |A_k|^2 \right)^{n/2}
            \biggr\rangle \ .
\end{equation}
Substituting this in Eq. (\ref{e2}), we arrive at the condition
\begin{equation}
\label{e15}
     \Bigl|\Bigl\langle \prod_{k=1}^n A_k \Bigr\rangle \Bigr| \le
          \frac{1}{n^{n/2}} \Bigl\langle
             \Bigl( \sum_{k=1}^n  A^\dagger_k A_k \Bigr)^{n/2} \Bigr\rangle ,
\end{equation}
which must be obeyed by a fully separable state.
Its violation yields the multipartite entanglement condition (\ref{cond2}).

An inspection of the conditions (\ref{cond1}) and (\ref{cond2}) 
reveals that for states such that
$A^\dagger_k A_k |\psi\rangle =A^\dagger_{k'} A_{k'} |\psi\rangle\
\forall k ,k' $, these conditions are the same.

In the case of  bipartite systems $n=2$, the inequality (\ref{cond1}) reduces to
\begin{equation}
\label{e15.1a}
   |\langle AB \rangle|^2 > \langle A^\dagger A\rangle
    \langle B^\dagger B\rangle,
\end{equation}
while the second inequality, Eq. (\ref{cond2}), becomes
$|\langle AB \rangle| > \frac{1}{2}
(\langle A^\dagger A \rangle + \langle B^\dagger B\rangle)$ or
\begin{equation}
\label{e15.1b}
    |\langle AB \rangle|^2 >  \langle A^\dagger A \rangle
         \langle B^\dagger B \rangle +
     \frac{1}{4}
          (\langle A^\dagger A \rangle-\langle B^\dagger B \rangle)^2.
\end{equation}
The bipartite entanglement condition (\ref{e15.1a}) is exactly one of those 
previously derived in Ref. \cite{Hillery06}, while the condition
(\ref{e15.1b}) is generally
weaker than the condition (\ref{e15.1a}) because the 
second term in the right-hand side is 
nonnegative.
However, 
without specifying the state of the system, there seems to be
no easy way to compare the two conditions for $n>2$. In fact, as we shall 
see in the next section, there are situations 
where the second condition, Eq. (\ref{cond2}),
can detect entanglement, while the first condition, Eq. (\ref{cond1}), cannot.

\section{Examples}
The two entanglement conditions presented in this paper can be applied to both discrete and
continuous systems.  We shall present examples of both.  These examples illustrate some of the
kinds of states for which these conditions can detect entanglement, and also the differences 
between the two conditions.

\subsection{GHZ-type states}

\subsubsection{Generalized GHZ state}

We begin by considering a system consisting of $n$ qubits in the state
\begin{equation}
\label{e20}
     |\psi\rangle = \cos\theta |0\rangle^{\otimes n}
              + \sin\theta |1\rangle^{\otimes n}\ .
\end{equation}
If we choose $A_k$ to be
\begin{equation}
\label{e21}
     A_k = |0\rangle_k\langle 1|,\quad
      A_k^\dagger A_k = |1\rangle_k\langle 1| ,
\end{equation}
then $A_k^\dagger A_k |\psi\rangle$ is independent of $k$, and, therefore, the
two conditions (\ref{cond1}) and (\ref{cond2}) are the same.
Using Eqs. (\ref{e20}) and (\ref{e21}), we find
\begin{eqnarray}
\label{e22}
      && \langle \psi| (\prod_{k=1}^n |0\rangle_k\langle 1|)
       |\psi\rangle  =
       \cos\theta \sin\theta,
\\
\label{e24}
       && \Bigl( \prod_{k=1}^n \langle \psi|
             \bigl( |1\rangle_k\langle 1| \bigr)^{n/2}
            |\psi \rangle \Bigr)^{1/n} =  \sin^2\theta.
\end{eqnarray}
If $|\cos\theta|>|\sin\theta|$, it can be seen that both entanglement 
conditions are satisfied, indicating the presence of entanglement.
Alternatively, one can choose
$     A_k = |1\rangle_k\langle 0|$, which implies that
     $ A_k^\dagger A_k = |0\rangle_k\langle 0| $. 
Then the left-hand sides of Eqs. (\ref{cond1}) and (\ref{cond2})
are unchanged and given again by Eq. (\ref{e22}), while the right-hand sides
become $\cos^2\theta$.  In this case,  entanglement is detected if
$|\sin\theta|>|\cos\theta|$. This choice of $A_k$ thus complements
the one given in Eq. (\ref{e21}). These two choices detect
entanglement in the state in Eq. (\ref{e20}) for all values of $\theta$, 
except for the case of $\cos\theta=\sin\theta$. 
For $\sin 2\theta \le 1/\sqrt{2^{n-1}}$ and $n$ odd, 
the state in Eq.\ (\ref{e20}) does not violate any $n$-party Bell inequalities
for correlation functions containing two dichotomic observables 
per local measurement station \cite{Zukowski02}, a set of inequalities that includes the 
Mermin-Klyshko inequalities \cite{Mermin90,Ardehali92,Klyshko}. 
Therefore, for this state the conditions (\ref{cond1}) and
(\ref{cond2}) are stronger criteria for entanglement detection than those coming from these Bell inequalities.
The entanglement in the state (\ref{e20}) also eludes detection by all
four spin squeezing inequalities derived in Refs. 
\cite{briegel1,briegel2}, which include those presented in
Refs. \cite{sorensen,korbicz,toth1}
as particular cases.

If the state has one spin flipped with respect to the rest
\begin{equation}
\label{e24.1}
     |\psi\rangle = \cos\theta |1\rangle \otimes|0\rangle^{\otimes (n-1)}
              + \sin\theta |0\rangle\otimes |1\rangle^{\otimes (n-1)}\ ,
\end{equation}
then by choosing
\begin{equation}
\label{e24.2}
     A_1 = |1\rangle_1\langle 0|,\quad
     A_k = |0\rangle_k\langle 1|,\quad k>1,
\end{equation}
one can readily find that entanglement is detected for
$|\cos\theta|>|\sin\theta|$. Generalization to cases where more
spins are flipped is straightforward.

Conditions (\ref{cond1}) and (\ref{cond2}) are also robust against 
noise. It is not difficult to verify that for the state
\begin{equation}
\label{e24.2a}
      \rho = p|\psi\rangle\langle \psi| + (1-p) 
         |0\rangle^{\otimes n} \langle 0|,
       \qquad 0<p<1,
\end{equation}
where $|\psi\rangle$ is given by Eq. (\ref{e20}),
they work the same as discussed above. Interestingly, this holds no matter 
how large the amount of noise is, that is how close $p$ is to zero, because
$p$ appears in the same way on both sides of the inequality and cancels out.
One can assume a more general type of noise
\begin{equation}
\label{e24.2b}
      \rho = p|\psi\rangle\langle \psi| + (1-p) 
         \frac{I}{2^n},
       \qquad 0<p<1,
\end{equation}
where $I$ is the unity operator.
With the choice of $A_k$ as in Eq. (\ref{e21}), condition
(\ref{cond1})  yields
\begin{equation}
\label{e24.2c}
        |\cos\theta \sin\theta| > \sin^2\theta + 
         \frac{1-p}{2p} .
\end{equation}
This inequality becomes increasingly difficult to satisfy as $p$ decreases, and impossible to
satisfy for $p \leq 1/3$.  Therefore, for states that are not too noisy, our conditions can still
detect entanglement.

\subsubsection{A partially separable state}

The state (\ref{e20}) is a genuinely multipartite entangled state.
We give now some examples to see how the two conditions (\ref{cond1}) and
(\ref{cond2}) work with a partially separable state.
Consider again an ensemble of $n$ spin $\frac{1}{2}$ particles, split
into two groups of $l$ and $(n-l)$ spins, each being in
a generalized GHZ state
\begin{equation}
\label{e24.3}
     |\psi\rangle = \bigl[ \cos\theta_1 |0\rangle^{\otimes l}
      + \sin\theta_1 |1\rangle^{\otimes l} \bigr]
     \otimes \bigl[ \cos\theta_2 |0\rangle^{\otimes (n-l)}
              + \sin\theta_2 |1\rangle^{\otimes (n-l)}\bigr]\ .
\end{equation}
Choosing $A_k$ as in Eq. (\ref{e21}), the inequalities in Eqs. (\ref{cond1})
and (\ref{cond2}) become
\begin{eqnarray}
\label{e24.4}
       |\cos\theta_1 \sin\theta_1 \cos\theta_2 \sin\theta_2|
        && > [(\sin\theta_1)^{2l} (\sin\theta_2)^{2(n-l)}]^{1/n},
\\
\label{e24.5}
      |\cos\theta_1 \sin\theta_1 \cos\theta_2 \sin\theta_2|
        && > \left(\frac{n-l}{n}\right)^{n/2}
              \cos^2\theta_1 \sin^2\theta_2
\nonumber\\
        && +\left(\frac{l}{n}\right)^{n/2} \cos^2\theta_2 \sin^2\theta_1
           + \sin^2\theta_1 \sin^2\theta_2,
\end{eqnarray}
respectively. Note that the two conditions now behave differently. 
It is apparent from the above equations that  entanglement cannot be detected if
$\sin\theta_1=0$ or $\cos\theta_1=0$, which we exclude from further
consideration. Since the inequalities (\ref{e24.4}) and (\ref{e24.5})
are rather involved, it is instructive to examine some special cases.

For $l=1$ and $n=3$, and $\sin\theta_1=\cos\theta_1=\frac{1}{\sqrt{2}}\ $, 
they become
\begin{eqnarray}
\label{e24.4a}
       |\cos\theta_2|
        && > (4 |\sin\theta_2|)^{1/3},
\\
\label{e24.5a}
      |\cos\theta_2 \sin\theta_2|
        && > 1.09 |\cos\theta_2 \sin\theta_2|
             +(1.24 |\sin\theta_2| - 0.44 |\cos\theta_2|)^2
              .
\end{eqnarray}
Obviously, when $\theta_2$ is close enough to $0$ or $\pi$, the first
 inequality, Eq. (\ref{e24.4a}), is satisfied meaning it can detect entanglement in the state,
while there exists no $\theta_2$ for which the second inequality,
Eq. (\ref{e24.5a}), is satisfied.

For $l=2$ and $n=4$, the two inequalities (\ref{e24.4}) and
(\ref{e24.5}) become
\begin{eqnarray}
\label{e24.4b}
       |\cos\theta_1 \sin\theta_1 \cos\theta_2 \sin\theta_2|
        && > |\sin\theta_1 \sin\theta_2|,
\\
\label{e24.5b}
      |\cos\theta_1 \sin\theta_1 \cos\theta_2 \sin\theta_2|
        && > \frac{1}{2}
    ( \cos\theta_1 \sin\theta_2 - \cos\theta_2 \sin\theta_1)^2
            +  2\sin^2\theta_1 \sin^2\theta_2,
\end{eqnarray}
respectively. It can be seen that
the inequality (\ref{e24.4b}) cannot be fulfilled for
any values of $\theta_1$ and $\theta_2$. 
Regarding the second inequality, we set for simplicity
$\cos\theta_1 \sin\theta_2 = \cos\theta_2 \sin\theta_1$,
to make it become
$\cos^2\theta_2 > 2 \sin^2\theta_2$, which clearly can be fulfilled
with $\theta_2$ in the neighborhood of $0$ and $\pi$. 
Thus the situation is
opposite to that occurring in the case of $l=1$ and $n=3$ discussed above in
that the second condition, not the first one, does better at detecting
entanglement.

In the limit of large $n$ but fixed $l$, Eqs.  (\ref{e24.4}) and (\ref{e24.5}) can be brought approximately to comparable forms
\begin{eqnarray}
\label{e24.6}
       &&|\cos\theta_2|> |\sin\theta_2|
         \frac{1}{|\cos\theta_1\sin\theta_1|},
\\
\label{e24.7}
       &&|\cos\theta_2|> |\sin\theta_2|
         \frac{1}{|\cos\theta_1\sin\theta_1|}
         \Bigl[ e^{-l/2} + \Bigl(1-e^{-l/2} \Bigr)
               \sin^2\theta_1 \Bigr],
\end{eqnarray}
where in going from Eq. (\ref{e24.4}) to Eq. (\ref{e24.6}), 
we made the
replacement $(\frac{\sin\theta_2}{\sin\theta_1})^{\frac{n-2l}{n}}
\rightarrow \frac{\sin\theta_2}{\sin\theta_1}$, while 
and in going from Eq.~(\ref{e24.5}) to Eq.~(\ref{e24.7}), 
we used the relation
$\left(\frac{n-l}{n}\right)^{n/2} = e^{-l/2}+O(\frac{1}{n})$
and dropped the vanishing second term in the right-hand side of Eq. (\ref{e24.5}).
Two comments can be made regarding the inequalities (\ref{e24.6}) and (\ref{e24.7}).
First, there exist $\theta_1$ and $\theta_2$ for which
both or one of them are satisfied, meaning entanglement is detected.
Second, since $\sin^2\theta_1<1$, the extra factor in
Eq.~(\ref{e24.7}) is less than unity with the result that the condition in
Eq.~(\ref{e24.7}) is more sensitive to entanglement than that in Eq.~(\ref{e24.6})
in the sense that there exist ranges of the parameters
$\theta_1$ and $\theta_2$
for which condition (\ref{e24.7}) can detect entanglement while
condition (\ref{e24.6}) cannot.

An estimate of how little entanglement in a multipartite system
is detectable by condition (\ref{cond1}) can be gained by studying
the state
\begin{equation}
\label{e24.8a}
     |\psi\rangle = \prod_{i=1}^{l\otimes} 
       \bigl[ \cos\theta_i |0\rangle_i
      + \sin\theta_i |1\rangle_i \bigr]
     \otimes \bigl[ \cos\theta |0\rangle^{\otimes (n-l)}
              + \sin\theta |1\rangle^{\otimes (n-l)}\bigr]\ ,
\end{equation}
where $l$ parties are separable, while the remaining parties
are in a GHZ state.
For this state, condition (\ref{cond1}) gives us
\begin{equation}
\label{e24.8b}
       |\cos\theta|> 
         \frac{1}{\prod_{i=1}^l|\cos\theta_i
             (\sin\theta_i)^{1-2/n}|}
        |\sin\theta|^{(n-2l)/n}.
\end{equation}
Since the denominator is less than one, the inequality can be satisfied 
when $l<n/2$. Though the number of 
separable 
parties has to be smaller than
half the total number of parties for entanglement to be detected,
using condition (\ref{cond1}) to look for 
entanglement is clearly less labor-intensive than using a bipartite 
condition to check every possible pairwise separation, especially in the case of large $n$.

\subsubsection{A mixed state}

Consider the $n$-party state
\begin{equation}
\label{e24.8}
     \rho = \frac{1}{n} \sum_{i=1}^n |\psi_i\rangle \langle \psi_i|,
\end{equation}
where
\begin{equation}
\label{e24.9}
     |\psi_i\rangle = \bigl[ \cos\theta_i |0\rangle
      + \sin\theta_i |1\rangle \bigr]
     \otimes \bigl[ \cos\theta |0\rangle^{\otimes (n-1)}_{\bar{i}}
              + \sin\theta |1\rangle^{\otimes (n-1)}_{\bar{i}}\bigr]\ ,
\end{equation}
$|0\rangle^{\otimes (n-1)}_{\bar{i}}$ and
$|1\rangle^{\otimes (n-1)}_{\bar{i}}$ being states where the spin $i$ is
excluded. $|\psi_i\rangle$ represents a state where the spin $i$ is
separated, while the remaining spins are in a generalized GHZ state. Though
$\rho$ is a statistical mixture of bipartite separable states, there is
no overall bipartite splitting with respect to which the state is separable.
Choosing $A_k$ as in Eq. (\ref{e21}), the inequalities
(\ref{cond1}) and (\ref{cond2}) are
\begin{eqnarray}
\label{e24.10}
      |\cos\theta\sin\theta \sum_{i=1}^n
            \cos\theta_i\sin\theta_i |
       &&>
        \Bigl[ \prod_{i=1}^n [\sin^2\theta_i + (n-1)\sin^2\theta]
                     \Bigr]^{1/n},
\\
\label{e24.11}
      |\cos\theta\sin\theta \sum_{i=1}^n
            \cos\theta_i\sin\theta_i |
        && > \left(\frac{n-1}{n}\right)^{n/2}
              \sin^2\theta \sum_{i=1}^n \cos^2\theta_i
\nonumber\\
        && +\left(\frac{1}{n}\right)^{n/2}
          \cos^2\theta \sum_{i=1}^n \sin^2\theta_i
           + \sin^2\theta \sum_{i=1}^n \sin^2\theta_i.
\end{eqnarray}
It is instructive to consider the case of very large $n$,
$\sin\theta_1=\cos\theta_1=\frac{1}{\sqrt{2}}$,
$\sin\theta_i=0$ for $i\ge 2$, 
for which these  inequalities simplify greatly to become
\begin{eqnarray}
\label{e24.12}
     && |\cos\theta|> 2(n-1) |\sin\theta|,
\\
\label{e24.13}
    && |\cos\theta|> \Bigl[ \frac{2}{\sqrt{e}} (n-\frac{1}{2}) +1\Bigr]
      |\sin\theta|,
\end{eqnarray}
respectively. It can be seen that both inequalities can be satisfied if
$|\cos\theta|$ is sufficiently close to one
and both would perform worse 
as 
the number of parties $n$ increases, the first more so than the second.

\subsection{Continuous-variable systems}

Consider an $n$-mode squeezed vacuum field state
\begin{equation}
\label{e25}
     |\psi\rangle = \sqrt{1-x^2} \sum_{m=0}^\infty
       x^m |m\rangle^{\otimes n},
\end{equation}
where $0< x< 1$. For the choice of $A_k$
\begin{equation}
\label{e26}
     A_k = a_k,\quad
      A_k^\dagger A_k = a^\dagger_k a_k ,
\end{equation}
$a_k$ being the annihilation operator of the field mode $k$,
the two conditions (\ref{cond1}) and (\ref{cond2}) are identical.
One finds that
\begin{eqnarray}
\label{e27}
       \langle \psi| (\prod_{k=1}^n a_k) |\psi\rangle
      = \frac{1}{x} (1-x^2)
         \sum_{m=0}^\infty x^{2m} m^{n/2},
\end{eqnarray}
\begin{eqnarray}
\label{e29}
        \Bigl( \prod_{k=1}^n \langle \psi|
             \bigl( a^\dagger_k a_k \bigr)^{n/2}
            |\psi \rangle \Bigr)^{1/n}
        = (1-x^2) \sum_{m=0}^\infty x^{2m} m^{n/2}.
\end{eqnarray}
A comparison of Eqs. (\ref{e27}) and
(\ref{e29}) shows that the inequalities
(\ref{cond1}) and (\ref{cond2}) are satisfied for any value of $x$
in the range $0<x<1$. That is to say, these conditions can always detect
entanglement in the multimode squeezed vacuum state.

We consider now an example of continuous variable systems where the two conditions (\ref{cond1}) and
(\ref{cond2}) work differently, namely a modified four-mode squeezed vacuum
state
\begin{equation}
\label{e30}
     |\psi\rangle = \sqrt{1-x^2} \sum_{m=0}^\infty x^m
        |m\rangle_1 |m\rangle_2 |m+1\rangle_3 |m+1\rangle_4,
\end{equation}
where $0< x< 1$. Choosing $A_k$ as in Eq. (\ref{e26}),
we find that
\begin{eqnarray}
\label{e35}
      && \langle \psi| (\prod_{k=1}^4 a_k) |\psi\rangle
      = \frac{2x}{(1-x^2)^2} \ ,
\\
\label{e36}
       && \Bigl( \prod_{k=1}^4 \langle \psi|
             \bigl( a^\dagger_k a_k \bigr)^2
            |\psi \rangle \Bigr)^{1/4}
      = \frac{x(1+x^2)}{(1-x^2)^2} \ ,
\\
\label{e37}
      && \frac{1}{16} \langle \psi|
             \Bigl( \sum_{k=1}^4 a^\dagger_k a_k  \Bigr)^2
            |\psi \rangle
        = \frac{1}{(1-x^2)^2}\frac{1}{4}(x^4+6x^2+1)\ ,
\end{eqnarray}
where we have made use of the relations
$       \sum_{m=0}^\infty x^{2m} = \frac{1}{1-x^2}$,
$ \sum_{m=0}^\infty x^{2m}m = \frac{x^2}{(1-x^2)^2}$, 
and
\mbox{$\sum_{m=0}^\infty x^{2m}m^2 = \frac{x^2(1+x^2)}{(1-x^2)^3}$}.
From Eqs. (\ref{e35}) and (\ref{e36}) it can be inferred that inequality
(\ref{cond1}) is satisfied for all $x$, meaning it always detects  entanglement. 
A comparison of Eq. (\ref{e35}) and (\ref{e37}) however shows that
inequality (\ref{cond2}) can be used for entanglement detection
 only for $x\stackrel{>}{\sim}0.1397$.
 
 
 \section{Conclusion}
 We have presented two sufficient conditions for determining when multipartite states are 
 entangled.  These conditions are quite flexible, because the operators appearing in them
 can be chosen to best match the systems being considered.  The conditions can be used to test
 for entanglement in discrete systems, continuous-variable systems, or mixtures of the two. 
 
 We have already seen that similar conditions for testing bipartite entanglement have proven
 useful in detecting entanglement in a variety of systems, including interacting spin systems and
 a collection of atoms interacting with the electromagnetic field.  We expect that the conditions
 derived here for multipartite systems will prove similarly useful.

\section*{Acknowledgments}
This research was supported by the National Science Foundation under grant PHY-0903660.

\end{document}